\pdfoutput=1

\documentclass[acmsmall]{acmart} 
\renewcommand\footnotetextcopyrightpermission[1]{} 
\AtBeginDocument{%
  }
    
\usepackage{algorithm}
\usepackage{algorithmic}

\usepackage{utfsym}

\usepackage{tabularx} 
\usepackage{array}

\usepackage{multirow} 
\usepackage{booktabs} 
\usepackage{makecell} 
\usepackage{array} 
\usepackage{multirow} 

\usepackage{ragged2e} 

\usepackage{caption}
\usepackage{array} 
\usepackage{cellspace} 
\usepackage{array} 
\usepackage{cellspace} 
\usepackage{colortbl} 
\definecolor{lightgray}{gray}{0.8} 

\usepackage[utf8]{inputenc}
\usepackage{listings}
\usepackage{xcolor}

\lstdefinelanguage{Solidity}{
    keywords={contract, function, public, payable, mapping, address, uint256, msg, sender}, 
    keywordstyle=\color{cyan!80!black}\bfseries, 
    morestring=[b]", 
    stringstyle=\color{orange!80!black}, 
    morecomment=[l]{//}, 
    morecomment=[s]{/*}{*/}, 
    commentstyle=\color{gray!60}\itshape, 
    sensitive=true 
}

\begin{document}
\bibliographystyle{plain}

\title{A Multi-Layered Security Analysis of Blockchain Systems: From Attack Vectors to Defense and System Hardening}


\author{Yuhuan Yang}
\authornotemark[1] 
\affiliation{%
  \institution{Hainan University}
  \city{Haikou}
  \country{China}}

\author{Shipeng Ye}
\authornote{These authors contributed equally to this work.}
\affiliation{%
  \institution{Hainan University}
  \city{Haikou}
  \country{China}}
\email{yeshipeng35@gmail.com}

\author{Xiaoqi Li}
\affiliation{%
  \institution{Hainan University}
  \city{Haikou}
  \country{China}}
\email{csxqli@ieee.org}

\renewcommand{\shortauthors}{Ye and Li}

\begin{abstract}
  The application of Bitcoin enables people to understand blockchain technology gradually. Bitcoin is a decentralized currency that does not rely on third-party credit institutions, and the core of Bitcoin's underlying technology is blockchain. With the increasing value of Bitcoin and the vigorous development of decentralization, people's research on blockchain is also increasing day by day. Today's blockchain technology has not only made great achievements in the application of Bitcoin, but has also been preliminarily applied in other fields, such as finance, medical treatment, the Internet of Things, and so on. However, with the initial application of blockchain technology on the Internet, the security of blockchain technology has also been widely concerned by people in the industry. For example, whether currency trading platforms, smart contracts, blockchain consensus mechanisms, and other technologies are vulnerable to attacks, and how we can defend against these attacks digitally and optimize the blockchain system is exactly the subject we want to study. For the security of appeal blockchain, this paper first analyzes the security threats faced by the application digital currency trading platform of the blockchain system, then analyzes the security problems of smart contract closely related to blockchain 2.0, and then analyzes and studies the security threats of blockchain public chain, consensus mechanism, and P2P. Finally, combined with the security problems at all levels of the blockchain system we analyze and study how to optimize the security of the blockchain system.
\end{abstract}



\maketitle
\fancyfoot{}
\pagestyle{plain} 

\section{INTRODUCTION}

The emergence of Bitcoin\cite{nakamoto2009bitcoin}has laid the foundation for the development of blockchain, and also indicates that blockchain technology is about to enter the peak of the tide of Internet development. With the application of blockchain technology on the Internet, such as Bitcoin, digital currency trading platform \cite{Regulatory_Rule_Zhang_Chao_2020}, Ethereum smart contract \cite{Ouyang_Liwei_2019_Smart_Contract}, etc. Whether these systems can run safely on the Internet has aroused people's attention. Even though blockchain technology is a brand new technology, it is deployed on the Internet after all, so it is inevitably exposed to various attack threats from the data link layer, network layer and application layer \cite{Yuquan_2020_from_Dengbao}. For example, in July 2017, Parity's multi signature wallet smart contract was attacked for the first time. In this attack, due to incorrect visibility settings of the contract, permission verification defects occurred, resulting in a loss of approximately \$30 million. On April 22, 2018, hackers launched an attack on BEC smart contracts, arbitrarily taking out BEC tokens and selling them in the market. BEC immediately depreciated sharply, with its value almost zero, and the market collapsed instantly. So, the direction I want to study is what attacks the blockchain system will face, how we should defend against these attacks, and how to optimize the system.

Regarding the security of smart contracts, Natoli et al.\cite{natoli2016blockchain} proposed the concept of blockchain anomalies, which cannot suspend transactions and may lead to double spending vulnerabilities, causing serious harm to smart contract users. In 2017, Atzei et al. \cite{atzei2017survey} first introduced the security vulnerabilities of the Ethereum platform and Solidity language, and then classified these vulnerabilities into three levels: Solidity, EVM, and Blockchain. Qiu Xinxin et al. \cite{Qiu_Xinxin's_2019_Ethereum_Smart_Contract} analyzed the re-entry vulnerability, integer overflow vulnerability, and denial of service vulnerability of smart contracts in China, and conducted detailed principle analysis and vulnerability reproduction for these common vulnerabilities, providing corresponding defense solutions. Zhao Hui, Li Xing, and others \cite{Zhao_Hui's_2021_Smart_Contract} analyzed the Delegatecall vulnerability and tx. origin vulnerability of smart contracts, and explained the security analysis tools available for smart contract vulnerabilities. Ni Yuandong, Zhang Chao, and others \cite{A_Review_of_Ni_Yuandong_2020} analyzed the vulnerabilities of smart contracts from the perspectives of a high-level language, Ethereum virtual machine, and blockchain. Research has also been conducted on attack scenarios such as double spending attacks\cite{karame2012double}and 51\% attacks \cite{budish2018economic} targeting blockchain consensus mechanisms both domestically and internationally. Attacks on P2P blockchain network technology have been studied mainly domestically and internationally, such as solar eclipse attacks \cite{heilman2015eclipse} and denial of service attacks \cite{elleithy2005denial}.

Section 2 first introduces blockchain, as well as its classification and development history; Section 3 analyzes the security threats of digital currency trading platforms from the perspectives of information collection, social engineering, input and output, server configuration security, and APP security; Section 4 analyzes vulnerabilities in Ethereum smart contracts, including integer overflow vulnerabilities, reentrancy vulnerabilities, denial of service vulnerabilities, short address vulnerabilities, and unverified return value vulnerabilities; Section 5 investigates the security of blockchain public chains, including double spending attacks based on public chain consensus mechanisms, as well as P2P based eclipse attacks and denial of service attacks; Section 6 will analyze how to optimize the blockchain system to address security threats from the data layer, network layer, consensus layer, contract layer, and application layer of the blockchain system; Section 7 provides a summary.

The main contributions of this paper are:
\begin{itemize}
\item \textbf{Analyze security threats:} A comprehensive analysis of blockchain security issues has been conducted at various levels, such as digital currency trading platforms, smart contracts, and public chains, covering multiple attack methods and types of vulnerabilities.
\item \textbf{Present defensive strategies:} In response to the above security threats, a series of specific defense measures are proposed, involving code optimization, transaction confirmation mechanism adjustment, network connection management, and other aspects.
\item \textbf{Optimize system security:} According to the network security level Protection 2.0 standard, from the data, network, consensus, and contract, application layer, the blockchain system security optimization scheme is presented.
\end{itemize}
This paper aims to comprehensively analyze the security situation of the blockchain system, find out the various attack threats it faces, explore effective defense strategies, and propose system optimization schemes to ensure the safe and stable operation of the blockchain system. At the same time, it also provides a reference for related research.

\section{Background}
\subsection{Blockchain}
I think we all started to understand the concept of blockchain from Bitcoin, so we often have the misconception that "Bitcoin is blockchain, and blockchain is also Bitcoin". Although Bitcoin and blockchain have a deep connection, this is only one specific application of blockchain, and there are many other applications of blockchain in society.

The concept of blockchain was first proposed by Satoshi Nakamoto in his 2008 publication "Bitcoin: A Peer to Peer Electronic Cash System", and he founded the Bitcoin network in 2009, developing the first block, the "Genesis Block". The Bitcoin network is a decentralized system where transactions between two nodes do not rely on third-party authorities, but operate through the consensus of each participant. Blockchain is a technology that originated from Bitcoin but is superior to Bitcoin.

The safe operation of blockchain can be attributed to the coordinated operation of four technologies. Distributed storage, P2P networks, consensus mechanisms, and cryptography are the four driving forces behind the development of blockchain technology\cite{hawashin2024blockchain}. Before the birth of blockchain, these four seemingly independent technologies combined to form blockchain technology. The electronic signature implemented using an elliptic curve encryption algorithm in cryptographic knowledge ensures the immutability of transactions, based on distributed storage to ensure the security and reliability of data, based on a P2P network to achieve synchronous updates and decentralization of distributed ledgers, and based on consensus mechanism to ensure that nodes in the blockchain system consciously handle things and operate.

\subsection{Classification and Characteristics of Blockchain Architectures}
Blockchain can be roughly divided into public chain, private chain, and consortium chain.

\begin{itemize}

\item Public chain is undoubtedly public, and users with needs can become nodes on the chain. Public chains use encryption technology to ensure that data is tamper proof, establish consensus in an untrusted environment, and form a decentralized trusted mechanism. The public chain is applicable to digital currency, e-commerce, Internet finance, and other application scenarios, such as Bitcoin and Ethereum.

\item Private chain is developed by a single entity. Private chains do not have the decentralization of blockchain, they are strongly centralized, and all permissions are owned by the chain owner. Private chains have the characteristics of high security and efficiency.

\item Consortium chain is a customized blockchain with an accounting system that provides services to specific organizations and individuals.

\end{itemize}

\subsection{Development History of Blockchain}
The development of blockchain has gone through three stages: blockchain 1.0, blockchain 2.0, and then blockchain 3.0, as shown in Fig 1\cite{Research_Progress_on_Attack_and_Defense_2021}.
\begin{figure}[h]
    \centering
    \includegraphics[width=1.0\textwidth]{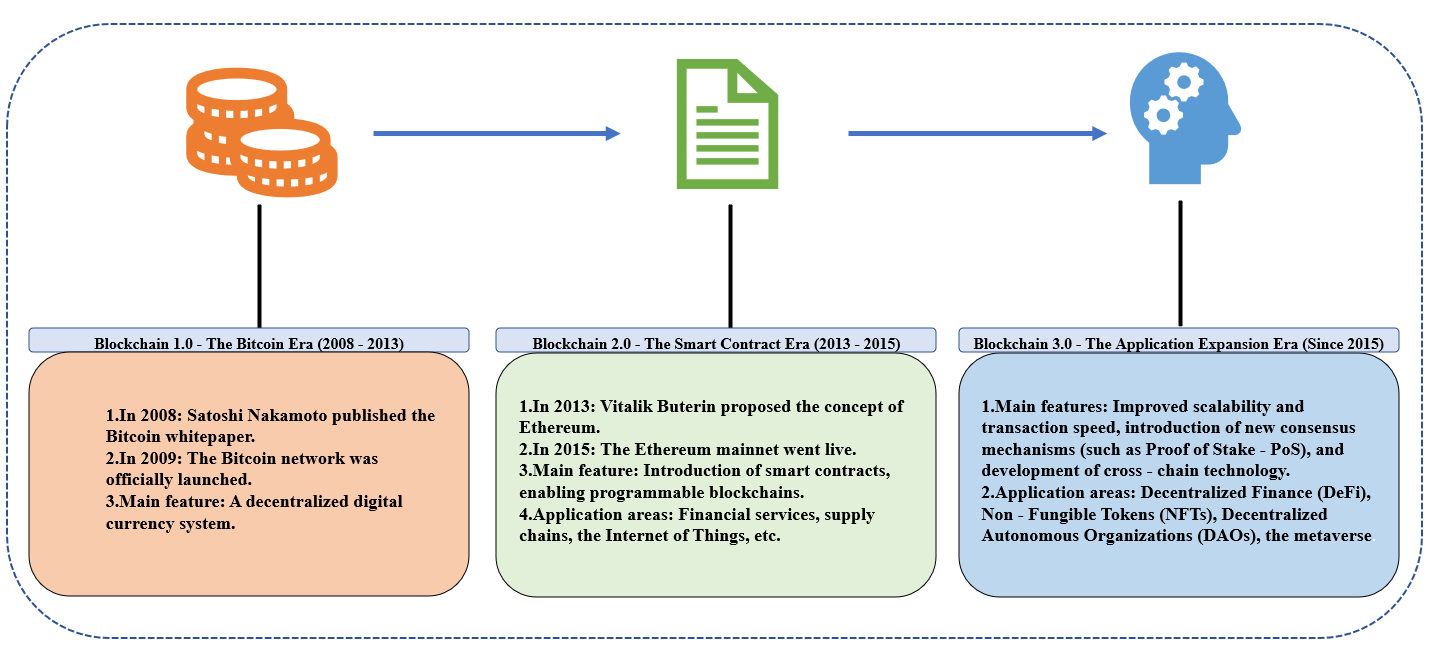}
    \caption{The Development History of Blockchain}
    \label{The Development History of Blockchain}
\end{figure}

The 1.0 stage of blockchain is represented by the Bitcoin system. With the rapid development of Bitcoin technology, people gradually understand blockchain technology as the underlying technology of Bitcoin.

After blockchain 1.0, the application scope of blockchain technology is no longer limited to encrypted digital currencies but can deploy smart contracts on the blockchain and open up decentralized applications. This is called the blockchain 2.0 stage, represented by Ethereum.

In the blockchain 2.0 stage, the use of smart contracts has further developed and applied blockchain technology\cite{li2021hybrid}, but its application scope is still relatively limited and lacks practical projects with practical value. With the development of blockchain technology, many organizations and enterprises have joined the ranks of researching blockchain applications. These organizations and enterprises have applied blockchain technology to solve many problems in the industry and improve the efficiency of business operations, which has entered the era of blockchain 3.0. At this stage, the industries involved in blockchain technology include virtualized assets, intelligent Internet of Things, supply chain management, decentralized operating systems\cite{li2024stateguard}, etc.

\section{SECURITY OF DIGITAL CURRENCY TRADING PLATFORMS}
The digital currency trading platform, also known as the digital currency exchange, is an important component of the blockchain industry, providing services for transactions between different digital currencies and between digital currencies and fiat currencies. It is also the main place for pricing and circulation of digital currencies\cite{niu2024unveiling,li2024defitail}. Digital currency flows in a peer-to-peer manner and can be decentralized, without the protection of third-party institutions or individuals. The application of distributed ledgers is a key innovation of digital currency, so digital currency is closely linked to blockchain technology.

\subsection{Threats to Digital Currency Trading Platforms}
A digital currency trading platform is a service platform provided by exchanges for customers\cite{kong2024characterizing}, which is equivalent to a C/S structure. Attacks on digital currency trading platforms can be considered from the perspective of our attacks on ordinary websites\cite{liu2025sok}. However, digital currency trading platforms have higher security and generally have fewer vulnerabilities such as ordinary SQL injection, cross site scripting attacks, client request forgery, and file uploads. So attacks on digital currency trading platforms need to be carried out from a more advanced level of penetration, such as collecting platform information first, combining social engineering, platform input and output, server configuration, and other aspects to launch attacks. Because the configuration information of these parts involves the infrastructure of digital currency trading platforms, this attack method is often more effective and requires a greater amount of work to protect, as it requires a thorough examination of the platform's basic information\cite{li2024guardians}.

\subsection{Information Gathering}
\subsubsection*{\textbf{Real IP Discovery of Server}}
Some large websites will activate CDN, and when customers located in different locations visit the website, they will first access the nearest CDN node server. Therefore, whether it is penetration testing or DDoS attacks, the target of the attack is the node server, which does not affect the security of the source server.

Finding the true IP address of the target is crucial when attacking the server. There are many ways to find the real IP address of a website, and the most common method is to obtain the server's real IP address by querying historical DNS records, and then bypass protection by scanning the port with the real IP address \cite{Wen_Chunsheng_2020_is_based_on}.

\subsubsection*{\textbf{Target Subdomain Detection}}
A website may have one main domain and multiple subdomains, and subdomain detection is an important component of information collection. Subdomain detection can help us discover more services in penetration testing, which increases the likelihood of discovering vulnerabilities. DNS domain transfer vulnerabilities can be exploited to detect subdomains by examining HTTPS certificates and performing enumeration mining.

\subsubsection*{\textbf{API Interface Information}}
API refers to calling functions written by others that can achieve specific functions. Due to the complexity of some computer operations, it is difficult for every programmer to manually implement them. Calling API can automatically implement these operations. Due to the great convenience brought by APIs, they are now widely used. Collecting API documentation and various parameters is very helpful for our penetration.

\subsubsection*{\textbf{GitHub Source Code Leak Discovered}}
Some developers are accustomed to uploading source code to code hosting platforms such as GitHub when writing code, and these source codes are what attackers want. If they obtain the source code, attackers can use code scanning tools to audit the code, which may scan for high-risk vulnerabilities in the source code.

\subsubsection*{\textbf{Whois}}
Although some exchanges now use the services provided by domain name registrants when registering domain names and do not disclose company or related personnel information on domain name registration websites such as Whois, there are still some exchanges that personally register domain names. In this case, using Whois or other tools can find detailed information about the registered company or related personnel of the exchange's domain name, which will be of great help for our future social engineering.

\subsubsection*{\textbf{Sensitive File Discovery}}
There are many types of sensitive files, among which the most classic and often most effective ones during testing are robots.txt, and sitemap. xml, and other files. Some files can even become a breakthrough point for attacks.

\textbf{Defense measures:} Information collection is the first step and a very important link in attacks.\cite{nguyen2024manipulating} Therefore, for digital currency exchanges, hiding system information is the first step to ensure their own security. Hiding system information well can resist most attacks.

\subsection{Social Engineering}
\subsubsection*{\textbf{Email Phishing and Website Phishing}}
When an attacker attacks an exchange, the first person they come into contact with is the customer service representative, making them the primary target of social engineering. Secondly, the responsible person we collect in information collection is also an important object of social engineering. They can construct a phishing email and send it to the customer service representative, the responsible person, and other email addresses. When a customer service representative or responsible person with low-security awareness clicks on the link, the attacker can obtain authentication from the log-in session, cookies, and other information.

Website phishing is when an attacker designs a phishing website in advance, and skilled attackers also operate the phishing website, making it indistinguishable from a normal website\cite{li2024detecting}. By inducing trading platform customer service personnel and responsible persons to log in to the phishing website, the attacker will obtain login account and password information, as shown in Fig 2.
\begin{figure}[h]
    \centering
    \includegraphics[width=1.0\textwidth]{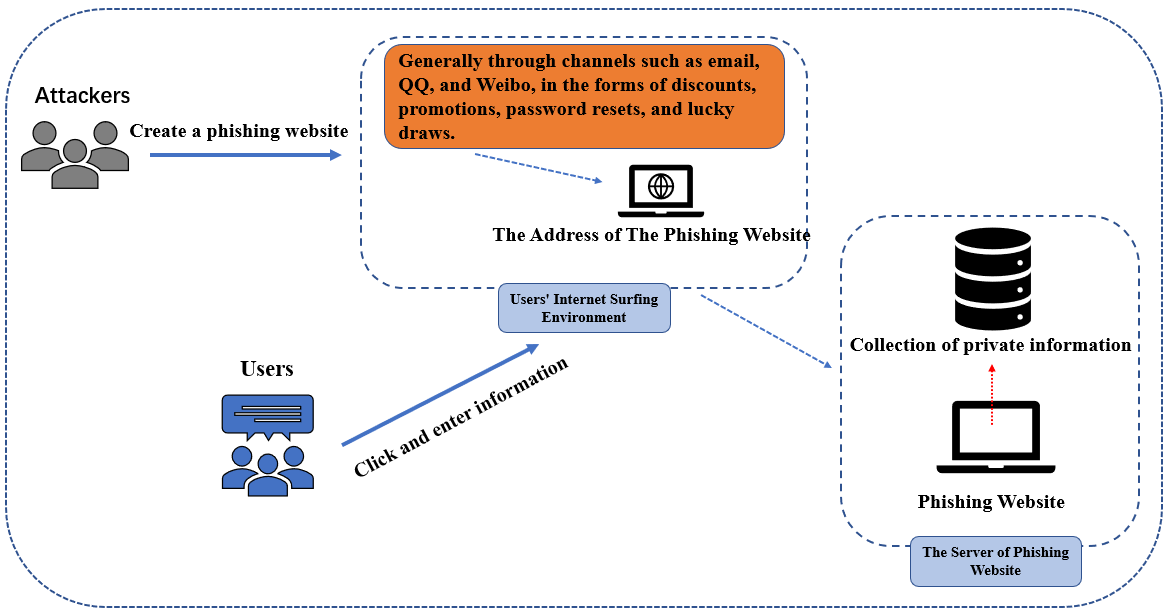}
    \caption{Website Phishing Attack}
    \label{Website Phishing Attack}
\end{figure}

\subsubsection*{\textbf{Harpoon Attack \cite{Chi_Yaping_2018_Spear_phishing_Attack}}}
Harpoon attack is a commonly used attack method in social engineering, which usually involves inserting malicious programs into email attachments and attaching a text that can lure the target to open the attachment. After the target opens the attachment, the malicious program will implant into the target's computer, allowing the attacker to continue infiltrating the internal network.

\textbf{Defense measures:} Social engineering undoubtedly targets people within the system, and there is no better way to prevent it. The only way to prevent social engineering is to provide security awareness training to system administrators and other personnel related to the system, enhance their security awareness, and not blindly open strangers' URLs, documents, and files. Only by having a certain understanding of some basic vulnerabilities can potential risks be avoided.

\subsection{Input and Output}
The security issues of input and output are often caused by developers' lack of security awareness during code development, who only focus on developing functional systems without achieving the goal of developing safe and reliable systems. And these input-output security issues are very serious for websites, posing a huge threat to databases, website management permissions, and internal networks. For example, file upload vulnerabilities can directly access the website's shell, command execution vulnerabilities can execute commands to bounce back the shell, and SQL injection vulnerabilities can view and modify database information.

\subsubsection*{\textbf{Cross Site Scripting Attack \cite{Pan_Jinkun's_2017_Cross_site}}}
Cross-site scripting attacks (XSS) are a common web vulnerability and the number one threat to client-side scripting security. XSS vulnerability submissions are also frequently seen on major trading platforms. The harm of XSS attacks is enormous and can be exploited in various ways, such as XSS phishing, cookie hijacking, and obtaining users' real IP addresses.

\subsubsection*{\textbf{SQL Injection Attack}}
SQL injection is very familiar to both security personnel and developers. SQL injection is when attackers input illegal data into the input interface provided by the server and obtain private data through output data, such as user accounts, passwords, etc.\cite{zhang2024sqlpsdem} It can be seen that the reason for SQL injection is that developers do not validate the validity of user input parameters when developing code, resulting in attackers constructing different SQL statements to perform arbitrary operations on the database. This shows the great harm of SQL statements.

\textbf{Defense measures:} For the defense against input-output attacks, the first step is to provide security training to code developers, especially during code development, third-party security review agencies should verify the security of the code.

\subsection{Server Configuration Security}
The server is dedicated to providing services for clients, usually with authentication and data transmission functions. Client login may require authorization from the server. If there are problems during the login process, attackers may enter the server, causing information leakage. In severe cases, it may lead to server damage and affect normal service provision. It can be seen that the security configuration of the server should not be taken lightly.

\subsubsection*{\textbf{Communication Protocol Security}}
The client needs a certain transmission protocol to establish a connection with the server, but some transmission protocols are very insecure. For example, the HTTP protocol is often exploited by attackers. Those who understand HTTP know that the HTTP protocol has GET and POST requests. If the client needs an account password to establish a connection with the server through the HTTP protocol, attackers can easily obtain the plaintext transmission account password from the HTTP GET packet using packet capture tools such as Wireshark.

\subsubsection*{\textbf{Manage Backend Security}}
The administrator of the management backend can perform many high-level operations such as managing and calling the resources of the entire system, so attackers will try their best to enter the management backend and gain access to it.\cite{hadi2024cost} It can be seen that the security of the management backend is very important. How to ensure the security of the backend? Firstly, it is necessary to hide the management backend, because the management backend is usually for relevant management personnel to log in, and the client does not need to log in to the management backend. Therefore, it is recommended not to expose the management backend to the public network or set a login whitelist, such as only allowing IP addresses of bastion machines to log in to the management backend.

\subsubsection*{\textbf{Weak Password and Default Password Detection}}
Weak passwords refer to passwords that are generally easy to guess, such as "123456", and "88888888", as well as birthdays, phone numbers, etc. The default password is the initial account password for some devices. Different manufacturers set their default login passwords when producing devices, which can generally be found on Baidu. So it is recommended that users change their default password and set a complex password after logging into the device.

\textbf{Defense measures:} Server security configuration is crucial, such as using secure communication protocols, hiding the management backend, and promptly modifying default login passwords.

\subsection{APP Security}
Today, with the continuous development of mobile devices, the frequency of people using mobile devices such as smartphones and tablets to access the internet is increasing, and the development of mobile apps is also becoming increasingly rapid\cite{li2017discovering}. In addition to being able to log in to the exchange through the webpage, users can also log in using the APP client.

\subsubsection*{\textbf{Code Decompilation Detection}}
If the APK file of the APP can be decompiled, it may lead to the leakage of source code information, and attackers can also conduct code audits to identify potential vulnerabilities\cite{sutter2024dynamic}. For such attacks, we should use security reinforcement tools to reinforce the APK files to prevent such attacks.

\subsubsection*{\textbf{Sensitive Information Leakage}}
Sensitive information leakage poses great risks, such as revealing test/administrator data, parameter annotation information, etc. You can check whether there is sensitive information leakage in the APP program files (such as source code, backup files, XML resource files, etc.).\cite{guo2025empirical} This type of vulnerability is easy to avoid, as long as developers raise their security awareness, most problems can be avoided.

\subsubsection*{\textbf{Data Storage Security}}
Check the storage directory of the APP. If the APP's storage files are globally readable and writable, then they can be operated by any user and modified and used by all applications. For such attacks, it is recommended not to store sensitive data of the app in /external storage. If necessary, these sensitive data should be encrypted.

\subsubsection*{\textbf{Log Information Leakage}}
In the development process of an app, to facilitate debugging, the log function is usually used to output some key process information, which often includes sensitive content such as execution processes,  user passwords in plain text, etc. This will allow attackers to better understand the internal structure of the app, conduct cracking and attacks, and even directly obtain valuable sensitive information. For such attacks, developers must remove log printing.

\subsubsection*{\textbf{Security Recommendations}}
At the vulnerability level, there are also OWASP10 vulnerabilities in the app, such as SQL injection, directory traversal, command execution, etc., but they are all caused by the lack of verification of data validity when external data is read. However, in addition to this, apps also have their own unique client security issues, which may pose risks such as app repackaging vulnerabilities, key leaks, HTTP packet capture, reverse engineering, etc. Nowadays, some manufacturers in the market have launched devices specifically designed for app vulnerability scanning. We can use these devices to regularly scan apps for vulnerabilities and prevent them from having high-risk vulnerabilities.

\section{ETHEREUM SMART CONTRACT SECURITY}

Smart contracts, literally speaking, are intelligent contracts that can be executed automatically\cite{medina2024blockchain}. The concept of smart contracts was first proposed by Nick Szabo in 1995, namely writing contracts in the form of code and automatically executing them without a third-party credit institution. Smart contracts can be enforced through technological means under certain conditions and contracts can be tracked and cannot be reversed.

Smart contracts were not utilized for a period of time after their proposal, as there was no system that could meet the requirements for the operation of smart contracts, until the emergence of blockchain technology and a decentralized trading system. In the era of blockchain 1.0, also known as Bitcoin, the blockchain system was not yet able to fully run Turing's smart contracts written in Solidity. In the era of blockchain 2.0, also known as Ethereum, Ethereum is an open-source platform where users can deploy smart contracts. Therefore, smart contracts have always been linked to blockchain\cite{li2025scalm}.

With the continuous development of blockchain technology, smart contracts have been widely applied in various industries, such as finance, gaming, the Internet of Things, etc. However,  security issues in smart contracts have also gradually become a concern. Smart contracts are typically used to manage digital assets, so attacks on smart contracts may result in a significant amount of assets for attackers\cite{li2024cobra,bu2025smartbugbert}. Starting from blockchain smart contracts, this section explores the types of attacks targeting smart contracts, as shown in Table 1, and how to defend against them.

\begin{table}[h]
\centering
\caption{The Main Attacks at Each Layer of Blockchain}
\begin{tabularx}{\linewidth}{|X|X|X|X|X|}
  \hline
  \rowcolor{lightgray}
  \textbf{Data Layer} & \textbf{Network Layer} & \textbf{Consensus Layer} & \textbf{Contract Layer} & \textbf{Application Layer} \\
  \hline
  \multicolumn{1}{|X|}{Collision Attack, Transaction Malleability Attack} &
  \multicolumn{1}{X|}{Eclipse Attack, Defer Bomb Attack} &
  \multicolumn{1}{X|}{Sybil Attack, 51\% Attack} &
  \multicolumn{1}{X|}{Reentrancy Attack, Integer Overflow Vulnerability, Resource-Exhaustion Vulnerability} &
  \multicolumn{1}{X|}{Selfish Mining Attack, Block-Withholding Attack} \\
  \hline
\end{tabularx}
\label{The Main Attacks at Each Layer of Blockchain}
\end{table}

\subsection{Integer Overflow Vulnerability}
Integer overflow is a common feature in computer programming languages, as programming languages have certain limitations on the storage space of data. Therefore, once the result of an integer operation exceeds this range, overflow will occur. Integer overflow vulnerabilities caused by arithmetic problems are not uncommon in programming, and in blockchain, Solidity, a commonly used language for writing smart contracts, also has integer overflow issues\cite{colin2024integrated}. There are three types of integer overflow in smart contracts: multiplication overflow, addition overflow, and subtraction overflow.

Addition overflow: For example, setting the max variable of type uint256 to its maximum value of 2 * * 256-1, then adding max+1 causes overflow, and the final output of max is 0, as shown in Listing 1.

\begin{lstlisting}[caption={Addition Overflow Vulnerability}]
    pragma solidity ^0.4.25;

    contract POC{
        function add_overflow()returns(uint256_overflow){
            uint256 max = 2**256- 1;
            return max + 1;
        }
    }
\end{lstlisting}

Subtraction overflow: Set the variable min of type uint256 to its minimum value of 0, let min -1, and the final value of min will become the maximum value of type uint256, which is 2 * * 256-1, as shown in Listing 2.

\begin{lstlisting}[caption={Subtraction Overflow Vulnerability}]
    function sub_underflow() returns (uint256_underflow){
        uint256 min = 0;
        return min - 1;
    }
\end{lstlisting}

Multiplication overflow: Set the variable mul of type uint256 to 2 * * 255, then multiply mul by 2 to become 2 * * 256. Exceeding the maximum value causes overflow, and the final return value is 0, as shown in Listing 3.
\begin{lstlisting}[caption={Multiplication Overflow Vulnerability}]
    function mul_underflow() returns (uint256_underflow){
        uint256 mul = 2**255;
        return mul * 2;
    }
\end{lstlisting}

\textbf{Defense measures:} Checking the user's input content can prevent integer overflow vulnerabilities, and the SafeMath function in the smart contract library can be used to encapsulate addition, subtraction, and multiplication interfaces, combined with the assert method for judgment. Integer overflow vulnerability is a code-level vulnerability. Although the SafeMath method can prevent the integer overflow vulnerability, if developers are unaware of the risks and forget to use this method, it can cause integer overflow. Therefore, to avoid such vulnerabilities, we also need to improve the security awareness of developers, and most importantly, undergo comprehensive security audits by the security team throughout the entire software development lifecycle.

\subsection{Reentrancy vulnerability}
Ethereum smart contracts can perform specific operations by calling external contracts\cite{wang2024smart}. If the external contract being called contains malicious code, the smart contract will be attacked by malicious code and cause huge losses(as shown in Fig 3). The most famous example is that hackers use reentrancy holes to attack DAOs and steal a large amount of Ethereum.
\begin{figure}[h]
    \centering
    \includegraphics[width=1.0\textwidth]{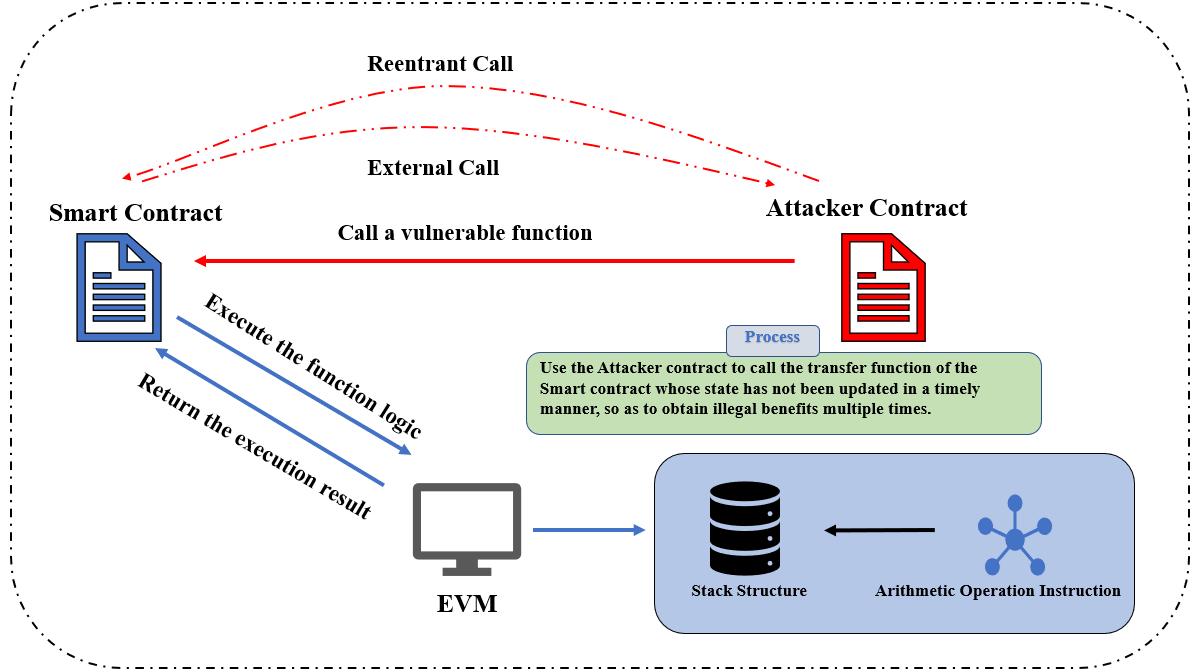}
    \caption{The Principle of Reentrancy Attack}
    \label{The Principle of Reentrancy Attack}
\end{figure}

Reentrancy vulnerability usually occurs when calling external contracts. The following code in Listing 4 analyzes how reentrancy vulnerability is exploited.
\begin{lstlisting}[caption={Contracts with Reentrancy Vulnerability}]
    pragma solidity ^0.4.19;
    contract A{
        mapping(address => uint256) balance;
        function deposit() payable public{
            balances[msg.sender] += msg.value;
        }
        function getMoney() public returns(uint256){
        return balances[msg.sender];
        }
        function getMoney(address add) returns(uint){
        return balances[add];
        }
        function withdraw(address add, uint amount){
        require(balances[add] > amount);
        add.call.value(amount)();
        balances[add] -= amount;
        }
    }
\end{lstlisting}

The functions of this contract mainly implement the user's deposit, record the user's deposit, and withdraw the user's withdrawal operations. There is no problem with the functionality of the contract implementation alone, but how does the reentrancy vulnerability attack the withdrawal function of this contract? From the perspective of the contract, the require function determines whether the user's assets are greater than the currently withdrawn coins\cite{li2024scla}. If they are, add.call.value will send the coins to the user's address and then subtract the withdrawn amount from the contract. Because of the call. value() function used here for currency transfer is different from the send() and transfer() functions, call. value() will use all the remaining gas for external calls (fallback function), while send() and transfer() only have 2300 gas to handle this operation. Therefore, the gas of send() and transfer() is not enough for external contracts to maliciously recursively transfer currency. In the contract, coins are first sent to the withdrawing user and then deducted from the asset. There is a recursion here. If the user uses a malicious recursive transfer contract in the called contract, all the money in the public wallet will be transferred out and stopped, causing huge losses.

\textbf{Defense measures:} The reentrancy vulnerability is the use of malicious recursion to achieve the goal of withdrawing all the money from the wallet\cite{liu2024gastrace}. We can defend against this vulnerability by adjusting the code to update the amount before the transfer, and then call it again. After understanding the differences between call. value(), send(), and transfer(), we can find that using send() or transfer() functions is safer when calling external contracts for transfers because they do not provide all the gas for use. Therefore, we should try to use send() and transfer() functions according to our functional requirements.

\subsection{Denial of Service Vulnerability}
We are all familiar with denial of service (DoS) attacks. In Ethereum smart contracts, attackers use DoS attacks to exhaust Ethereum smart contract and system resources, making it impossible to provide normal services to the outside world\cite{chen2020survey}. Although Ethereum smart contracts have a gas mechanism in place to prevent attackers from causing the smart contract to crash by consuming system and network resources, attackers can continue to execute high consumption code in the smart contract, preventing EVM from providing normal services to legitimate users\cite{Zhao_Gansen's_2019_Smart_Contract}.

Denial of service attacks can be divided into three categories: progress state based on external calls, manipulation of mappings or array loops through external means, and owner actions.
\begin{lstlisting}[caption={Code with Denial of Service Vulnerability}]
    pragma solidity ^0.4.10;

    contract presidentofCountry{
        address public president;
        uint256 price;
        function presidentofCountry(uint256 _price) {
            require(_price > 0);
            price = _price;
            president = msg.sender;
        }
        function becompresident() payable {
            require(msg.value >= price);
            president.transfer(price);
            president = msg.sender;
        price = price * 2;
        }
    }
\end{lstlisting}

Listing 5 is a denial of service code based on external calls. In the becomepresident function of the contract code, require(msg.value >= price); Determine that the input Ethereum is greater than the price, president.transfer(price); Return the Ethereum in the price to the input user, presidengt = msg.sender; Set the price to twice its original value.

The encoding method and the logical relationship of some smart contracts can easily be exploited by attackers to cause DoS attacks, such as causing the smart contract to report errors repeatedly and being unable to provide services to normal users.

As shown in Listing 6 based on the DoS attack contract, if the account that returns the transfer amount is the attacking contract's account, and when the attacking contract becomes the new president, write the fallback function in the contract and write revert() in the fallback function to run an error, the smart contract will continue to loop through errors, preventing other accounts from becoming the new president and causing the DoS attack.

\textbf{Defense measures:} The DoS attacks listed above are based on external calls to smart contracts. To prevent such attacks, smart contract developers need to think carefully about the possible problems caused by external calls before making corresponding assignment calls to avoid this type of loop error that can cause the smart contract to crash.
\begin{lstlisting}[caption={DoS Attack Code}]
    contract attack {
        function () {
        revert();
        }
        function attack(address _target) payable {
        bytes4 funcSelector = bytes4(keccak256("becompresident()"));
        _target.call.value(msg.value)(funcSelector);
        }
    }
\end{lstlisting}

\subsection{Short Address Vulnerability}
The short address vulnerability is caused by the fact that smart contracts do not validate the validity of the address length entered by users, so some malicious users create short address vulnerabilities by entering addresses shorter than normal addresses\cite{javanmardi2024m}. The principle of the short address vulnerability is that if the address being transferred is 32 bytes, it can be referred to as 0000000 0000000 00001234567890098765432112345678900987654321. The transfer amount is 32 bytes. Now the attacker intentionally removes the last two digits of the transfer address, and the 00 in front of the transfer address will be added to the end of the transfer address, resulting in a missing 00 after the transfer amount. Therefore, when the data enters EVM, the mechanism of EVM will automatically add 00 after the transfer amount, and the transfer amount will become 0000000 0000000 0000000 0000000 0000000 0000000 000000 00000 200, changing from transferring 2 Ethereum to transferring 512 Ethereum. This is the exploitation of the short address vulnerability.

\textbf{Defense measures:} In fact, the principle of short address vulnerabilities is very simple, but sometimes they are overlooked by developers. When developing smart contracts, as long as we use whitelists to verify the validity of user input data, we can avoid short address vulnerabilities.

\subsection{Unverified Return Value Vulnerability}
When making external calls, if the return value is not checked, it can easily lead to vulnerabilities\cite{sun2024gptscan}. Smart contracts have three underlying calls, call()delegatecall()callcode(), And three conversion functions, call.value()send()transfer().

Call() is used in Solidity to make external calls and returns a boolean value to indicate whether the call was successful. Both Delegatecall() and call() return a boolean value, but the difference is that call() returns the current contract after the external call is completed, while delegatecall() copies the external code into the current contract for execution. Callcode() is similar to delegatecall(), except for the difference in pointing to msg.sender and msg.value. The call to delegatecall() does not modify the content of msg.

The functions and differences of Call.value(), send(), and transfer() are introduced in Section 3.3.
\begin{lstlisting}[caption={Unverified Return Value Vulnerability Code}]
   function withdraw(uint256 _amount) public {
        require(balances[msg.sender] >= _amount);
        balances[msg.sender] -= _amount;
        etherLeft -= _amount;
        msg.sender.send(_amount);
    }
\end{lstlisting}

As shown in the code in Listing 7, the return value of send() is not checked in msg. sender. send (a\_mount). After the fallback() function call fails, send() will return false. The account balance will decrease, but the transfer will not be successful.

\textbf{Defense measures:} The unchecked return value vulnerability is a code level vulnerability, so developers should try to use the transfer() function instead of the send() function during development. If the send() function is to be used, the return value must be checked, and in the development of smart contracts, it must be reviewed by third-party security review agencies. A better approach is to adopt the withdrawal mode, in which each user needs to call an independent function to handle the issue of sending Ethereum from the contract, allowing users to independently handle the results of sending transactions.

\section{PUBLIC CHAIN SECURITY}
We usually divide blockchain into public chain, consortium chain, and private chain\cite{zhang2024blockchain}. In fact, there is no essential difference between the three types of blockchain. For example, a consortium chain is built for specific groups of people to use. However, when the number of users of consortium chains continues to expand and reaches a certain level, in a sense, there is no difference between public chains and private chains. Therefore, we will study the most representative public chain security.

\subsection{Security of Consensus Mechanism}
The consensus mechanism is the soul of the blockchain technology system and the key to ensuring that all nodes in the network reach the correct consensus. Below, we will study the security threats that the PoW consensus mechanism used in Bitcoin and Ethereum is vulnerable to.

\subsubsection*{\textbf{PoW consensus mechanism}}
The nodes of Bitcoin and Ethereum compete for accounting rights through computing power, and the process of competing for accounting rights is "mining". Bitcoin and Ethereum use the PoW mechanism to determine who has this accounting right\cite{song2024unveiling}. There are three principles to follow when mining through the PoW mechanism: only one person can successfully account for a period of time; Compete for unique accounting rights by solving cryptographic problems; Verify other nodes and copy the accounting results. After understanding the principle of PoW mechanism, one can start researching attacks against PoW mechanism.

\subsubsection*{\textbf{Double Spending Attack}}
Double spending attack is the most classic attack in Bitcoin and Ethereum. Double spending attack refers to the attacker being able to use the same token to complete multiple payments\cite{sun2024doubleup}. The analogy is that we spend 100 yuan but spend twice, buying something worth 200 yuan. There are several ways to launch a dual flower attack: 51\% attack, racial attack, and Fanny attack\cite{Overview_of_Typical_Security_Issues_2021}.

\subsubsection*{\textbf{51\% Attack}}
The 51\% attack refers to an attacker controlling over 50\% of the computing power on the entire network\cite{saveetha2024integrated}. The PoW mechanism allows multiple sub chains to exist simultaneously, but the blockchain only recognizes the longest chain as the main chain with the highest workload. Here is an example of how a 51\% attack is implemented. If the attacker broadcasts the same transaction to two different blockchain networks, and two miners in both networks almost simultaneously obtain accounting rights.

If a miner chooses to continue accounting on Block 4, and the merchant only needs two confirmations to complete the transaction, then Block 4 will become the main chain, and the transaction will be completed.

After the merchant confirms the transaction, the attacker immediately becomes a miner because they control 51\% of the network's computing power. Therefore, the probability of the attacker obtaining accounting rights greatly increases. If the attacker obtains accounting rights twice in a row on the Block5 chain, then Block5 will become the main chain, and transactions on the Block4 chain will be traced back. The virtual currency will return to the attacker's operation, and the double spending attack will be completed, as shown in Fig 4.
\begin{figure}[h]
    \centering
    \includegraphics[width=1.0\textwidth]{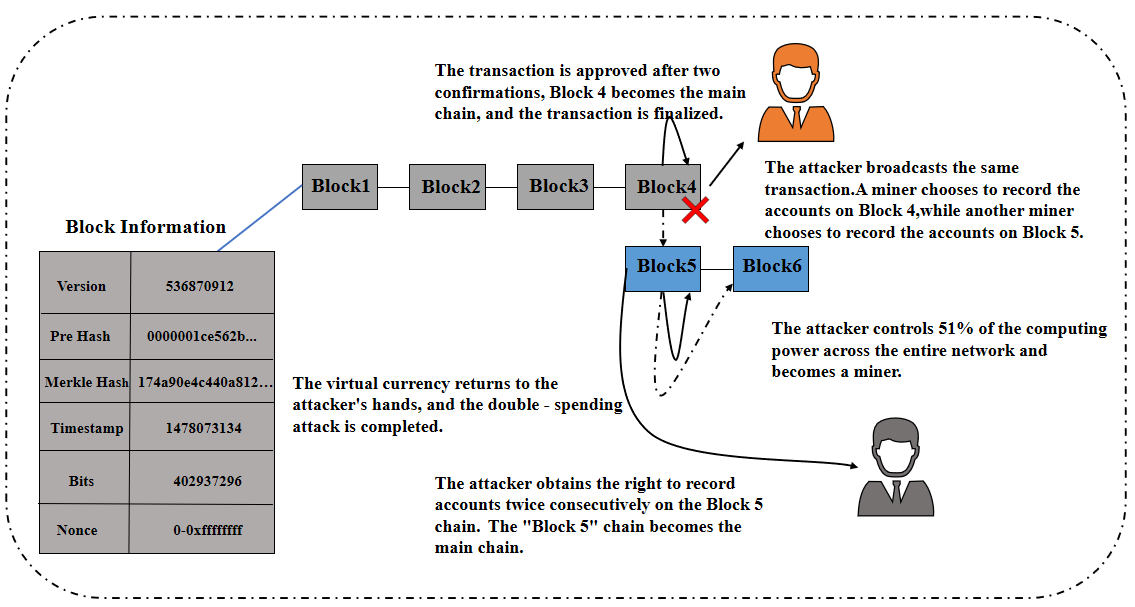}
    \caption{The double-spending attack caused by the attacker controlling 51\% of the computing power}
    \label{51_Attack}
\end{figure}

\textbf{Defense measures:} Blockchain node clients need to enhance their host's ability to defend against viruses and Trojans, preventing them from becoming zombie hosts and being controlled by attackers to steal mining resources. We, as merchants, can increase the number of confirmations for successful transactions. In the above example, only two confirmations are needed to complete the transaction successfully. However, increasing the number of confirmations to 100 can reduce the success rate of the attack, but the disadvantage of doing so is that the transaction latency is relatively high.

\subsubsection*{\textbf{Race Attack}}
Race attacks mainly achieve double spending by controlling miner fees. For example, if an attacker sends a certain number of tokens to a merchant, we name it branch A. If the merchant implements 0 confirmation, the attacker will then send this token to their own wallet, which we will name branch B. A Race attack is when the attacker increases the miner fee on branch B, so branch B is likely to be packaged by miners before branch A\cite{tyagi2024role}. If the branch that the attacker sent to their wallet is pre-packaged, the token on branch A will roll back, achieving a double spending attack.

\textbf{Defense measures:} From the principle of race attacks, businesses can set a higher number of confirmations to defend against such attacks.

\subsubsection*{\textbf{Finney attack}}
Attackers can achieve double spending by controlling the broadcast time of blocks, which is known as the Finney attack\cite{masteika2024bitcoin}. For example, if an attacker transfers money from address A to address B, and both addresses are the attacker's own, the transaction will not be broadcasted for now. At this point, the attacker transfers the transaction to the merchant's C address using address A, and the merchant defaults to 0 confirmation to complete the transaction. The attacker also takes away the goods. At this point, the attacker broadcasts the transaction of A transferring money to B. Since the transaction sent to their own address precedes the transaction sent to the merchant, by controlling the broadcast time, the attacker completes a double spend.

\textbf{Defense measures:} The most effective way to defend against such attacks is for merchants to prohibit 0-confirmation transactions and set a certain number of confirmations.

\subsection{P2P Security}
Blockchain is a decentralized database that requires nodes in the network to exchange information and reach consensus\cite{li2024coraldb}. The blockchain adopts a peer-to-peer data exchange network architecture, which has higher network security, but there are also some security threats \cite{Research_on_the_Security_2020_Ma_Zhuoyuan}.

\subsubsection*{\textbf{Eclipse Attack}}
In the Bitcoin network, each node can create 8 external TCP connections and accept 117 internal TCP connections, and all connections will be reset when the node restarts. Based on this principle, Hlianman et al. proposed a solar eclipse attack on blockchain in 2015. Each node will select the host IP address to establish a connection from the new and tried tables, and will prioritize the IP with the closest timestamp\cite{shi2025eclipse}. The attacker will try every means to make the target node restart continuously, update the new and tried tables of the target node, and fill these two tables with the attacker's IP, so that the attack target can only connect to the attacker's IP and cannot connect to other legitimate nodes, resulting in the attacked node being unable to maintain the ledger together with other legitimate nodes \cite{Research_Progress_on_Attack_and_Defense_2021}.

\textbf{Defense measures:} As mentioned above, attackers using a solar eclipse attack need to establish a connection with the target node, so they should try to bury their attack IP in the new and tried tables as much as possible and repeat the connection to prioritize the timestamp of the attack IP, thereby increasing the connection probability. In response to a solar eclipse attack, we can randomly choose IP addresses instead of prioritizing the old and new timestamps. We can also prohibit direct access to remote new and tried tables, making it difficult for attackers to bury these two tables.

\subsubsection*{\textbf{Denial of Service Attack}}
The attack utilizes a large number of zombie hosts in the network to attack the target node, attacking the system or network resources of the target node, causing the target system to crash and thus deny service, as shown in Fig 5. This is a typical attack method in computer networks. Although it consumes a lot of resources from attackers, failing to prevent such attacks can cause serious damage to our services\cite{Tangsha's_2020_Blockchain_Technology}.

\textbf{Defense measures:} In response to malicious resource-occupying attacks such as DoS and DDoS, we can deploy security devices, such as protective walls that can choose to enable DoS and DDoS defense.
\begin{figure}[h]
    \centering
    \includegraphics[width=1.0\textwidth]{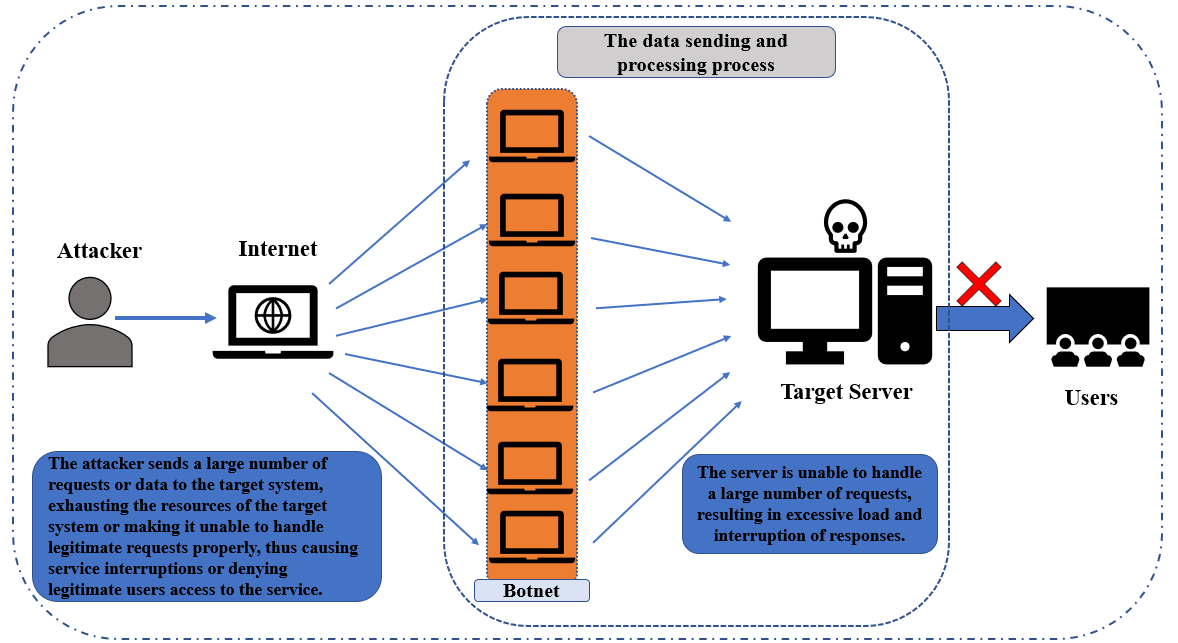}
    \caption{The Principle of DoS Attack}
    \label{DoS Attack}
\end{figure}

\section{SYSTEM SECURITY OPTIMIZATION}
The research on blockchain system security is a holistic level that addresses security issues at the application layer, contract layer, consensus layer, network layer, and data layer of the blockchain system. It is planned, supervised, and optimized using the Network Security Level Protection 2.0 standard \cite{Zhuyan_2020_Blockchain,Yuquan_2020_from_Dengbao,Yu_Luting_2021,Research_on_the_Security_2020_Ma_Zhuoyuan}.

Data layer: Encrypt sensitive information using nationally certified cryptographic techniques, such as elliptic curve public key cryptography (SM2), digest algorithm (SM3), etc; Using asymmetric encryption algorithms for information encryption, digital signatures, login verification, and other scenarios; We need to establish a comprehensive management system for the keys of the blockchain system and register the users of the keys\cite{perez2024blockchain}; Regularly update the encryption algorithms used in the blockchain system, eliminate those that have already been compromised, and use more secure new encryption algorithms;

Network layer: Level protection evaluation. Common system evaluations in our daily lives include secure communication networks (network architecture), secure area boundaries, and secure computing environments. However, the blockchain system does not have a central server and is maintained and updated through various legitimate nodes\cite{yang2024blockchain}. Therefore, trusted verification technology must be used to verify the trustworthiness of nodes accessing the blockchain. When a node accesses the system, its identity is verified, and only legitimate nodes can enter the blockchain system; Network jitter will not have a significant impact on data transmission, and the system can operate stably; Check network configuration, set a session no response time, and be able to end sessions that do not respond within a limited time to prevent ineffective use of system resources; Like ordinary systems, blockchain systems are created to meet certain customer needs, so it is necessary to ensure that the bandwidth of each part of the blockchain network meets the peak business demand and that the business processing capacity of nodes meets the peak demand; When transmitting data at the network layer, using encryption transmission protocols such as SSH and HTTPS can ensure that the data is not tampered with or eavesdropped upon during point-to-point transmission, broadcast communication, and forwarding. This can prevent the key from being obtained by illegal users through packet capture during transmission at the network layer; The blockchain system should have logs that record node status information, which can provide reliable node data for the stable operation of the system;

Consensus and Contract Layer: Regularly monitor the usage of resources in the computer, and the consensus mechanism should adhere to the principle of minimizing the consumption of computer resources; Real-time backup of transaction data generated in the blockchain by each node; The blockchain system should set a consensus threshold so that nodes exceeding the threshold reach consensus, which represents consensus among all nodes in the network and can prevent the influence of abnormal nodes\cite{venkatesan2024blockchain}; By verifying the correctness and logic of transactions, the cost of malicious fraud can be increased to punish it and avoid malicious consensus; Hand over the developed smart contract to a third-party authoritative organization for professional testing to prevent code vulnerabilities or logical flaws in the smart contract; Train code developers and require them to develop contract code in accordance with established development standards; Be cautious when using external contract calls to prevent risks caused by malicious contracts;

Application layer: Digital currency trading platforms, should undergo strict evaluation by evaluation institutions and be rectified and constructed according to the requirements of level protection; Security service companies can also conduct penetration testing to uncover deeper vulnerabilities in real attack and defense environments.

\section{CONCLUSION}
Blockchain technology is constantly developing, and security is the foundation for the stable development of blockchain systems. To address attacks at different levels of blockchain, we need to have targeted defenses and the ability to plan the overall security of the system. According to the national network security level protection, continuously improve the security of the blockchain system, perfect the system management system, and promote the safe and healthy development of blockchain applications.

\bibliography{samples/main}
\end{document}